# Piezoelectric nonlinearity and frequency dispersion of the direct piezoelectric response of BiFeO$_3$ ceramics

Tadej Rojac,[1,a)] Andreja Bencan[1], Goran Drazic[1], Marija Kosec,[1] and Dragan Damjanovic[2]

[1]Jožef Stefan Institute, Jamova cesta 39, 1000 Ljubljana, Slovenia

[2]Swiss Federal Institute of Technology – EPFL, Ceramics Laboratory, Lausanne 1015, Switzerland

**Abstract**

We report on the frequency and stress dependence of the direct piezoelectric d$_{33}$ coefficient in BiFeO$_3$ ceramics. The measurements reveal considerable piezoelectric nonlinearity, i.e., dependence of d$_{33}$ on the amplitude of the dynamic stress. The nonlinear response suggests a large irreversible contribution of non-180° domain walls to the piezoelectric response of the ferrite, which, at present measurement conditions, reached a maximum of 38% of the total measured d$_{33}$. In agreement with this interpretation, both types of non-180° domain walls, characteristic for the rhombohedral BiFeO$_3$, i.e., 71° and 109°, were identified in the poled ceramics using transmission electron microscopy (TEM). In support to the link between nonlinearity and non-180° domain wall contribution, we found a correlation between nonlinearity and processes leading to deppining of domain walls from defects, such as quenching from above the Curie temperature and high-temperature sintering. In addition, the nonlinear piezoelectric response of BiFeO$_3$ showed a frequency dependence that is qualitatively different from that measured in other nonlinear ferroelectric ceramics, such as "soft" (donor-doped) Pb(Zr,Ti)$O_3$ (PZT), i.e., in the case of the BiFeO$_3$ large nonlinearities were observed only at low field frequencies (<0.1 Hz); possible origins of this dispersion are discussed. Finally, we show that, once released from pinning centers, the domain walls can contribute extensively to the electromechanical response of BiFeO$_3$; in fact, the extrinsic domain-wall contribution is relatively as large as in Pb-based ferroelectric ceramics with morphotropic phase boundary (MPB) composition, such as PZT. This finding might be important in the search of new lead-free MPB compositions based on BiFeO$_3$ as it suggests that such compositions might also exhibit large extrinsic domain-wall contribution to the piezoelectric response.

[a)]Electronic mail: tadej.rojac@ijs.si

## I. INTRODUCTION

Bismuth ferrite (BiFeO$_3$) possesses a variety of interesting properties, including magnetoelectric coupling, high intrinsic polarization (100 μC/cm$^2$ in the (111)$_{pc}$ direction), and high Neel (T$_N$ = 370°C) and Curie (T$_C$ = 825°C) temperatures.[1–3] Owing to its rhombohedral structure, the ferrite has been considered as an end-member of a variety of lead-free solid solutions exhibiting morphotropic phase boundary (MPB), which is believed to be the key for obtaining high piezoelectric properties. Some recently studied systems are BiFeO$_3$–REFeO$_3$ (RE = La, Nd, Sm, Gd, Dy),[4–6] BiFeO$_3$–BaTiO$_3$,[7,8] and BiFeO$_3$–Bi(Zn$_{0.5}$Ti$_{0.5}$)O$_3$.[9] In addition, the high T$_C$ of BiFeO$_3$ enables to design MPB compositions suitable for high-temperature piezoelectric applications, such as the recently emerged BiFeO$_3$–PbTiO$_3$.[10]

In spite of the extensive research on BiFeO$_3$ and its chemical modifications, the knowledge of the piezoelectric response of pure BiFeO$_3$ ceramics is rather poor and usually only values of d$_{33}$ are reported, which range from 4 to 60 pm/V.[11–14] The reason for the large spread in the values is probably a combination of high electrical conductivity and high coercive field (50–85 kV/cm), which are strongly processing sensitive,[11,15–18] and difficulties in processing the ferrite.[19–21] The problem of the high electrical conductivity, which prevents application of high electric fields to BiFeO$_3$, is often overcome by measuring the electromechanical response only locally using piezoforce microscopy (PFM).[9,12,22]

Using a mechanochemically assisted synthesis, we have recently succeeded to prepare BiFeO$_3$ ceramics with sufficiently low DC conductivity to withstand large electric fields, i.e., up to 180 kV/cm.[15] As a result of electric-field switching of domains, a large strain, comparable to that achievable in Pb-based ferroelectric ceramics with MPB compositions, such as Pb(Zr,Ti)$O_3$ (PZT) and Pb(Mg,Nb)$O_3$–PbTi$O_3$ (PMN-PT), was measured in these BiFe$O_3$ ceramics.[23] Those results suggested that the weak-field electromechanical response of the ferrite may also involve a considerable contribution of non-180° domain-wall movement. In general, the motion of non-180° domain walls under subcoercive fields is considered as the most important extrinsic contribution to the piezoelectric properties of ferroelectric ceramics and may reach up to 60-70% of the total piezoelectric response.[24–26] Thus, besides large switchable strain, BiFe$O_3$ ceramics are expected to exhibit relatively large weak-field piezoelectric response.

Dielectric nonlinearity, i.e., dependence of the permittivity on the electric field amplitude, which is usually attributed to the irreversible domain wall movement, has recently been investigated in BiFe$O_3$ thin films.[27] It was shown that the dielectric permittivity can increase up to 40% of its initial low-field value as the amplitude of the AC driving field is increased; this nonlinear contribution depended strongly upon the domain structure of the film. Another example of domain-wall contribution to properties was found in donor-doped BiFe$O_3$. It was shown that doping BiFe$O_3$ with W$O_3$ increased the piezoelectric $d_{33}$ coefficient of the films by 60%, presumably due to a decreased concentration of oxygen vacancies and consequent reduced pinning of domain walls.[28] In agreement with our earlier study of the strain–electric-field response,[23] all these results suggest a potentially large nonlinear contribution in BiFe$O_3$ provided by domain-wall motion under subcoercive fields.

We report here a study of the direct piezoelectric response of BiFe$O_3$ ceramics. Dependence of the piezoelectric $d_{33}$ coefficient on frequency, and dynamic (alternating or "AC") and static (or "DC") stress are presented. The relationship between domain-wall contributions and piezoelectric nonlinearity was studied by means of methods that provide depinning of domain walls from defects, i.e., quenching and high-temperature sintering. In addition, we performed a domain structure study using TEM. We show that the piezoelectric response of $BiFeO_3$ is strictly different from that in other perovskite materials reported so far, including "hard" and "soft" PZT, $BaTiO_3$, $Bi_4Ti_3O_{12}$, etc.

## II. EXPERIMENTAL PROCEDURE

$BiFeO_3$ ceramics were prepared by mechanochemical activation of a $Bi_2O_3$–$Fe_2O_3$ powder mixture followed by direct sintering at 760°C for 6 h or 880°C for 10 h. Details of the preparation procedure are given in ref. 15. The geometrical relative densities of the ceramics sintered at 760°C and 880°C were 92% and 95%, respectively.

The phase composition of the sintered samples was determined by X-ray diffraction (XRD) analysis using a Panalytical X'Pert Pro diffractometer. The concentration of the secondary phases ($Bi_{25}FeO_{39}$ and $Bi_2Fe_4O_9$) was determined by Rietveld refinement method using Topas software package[29] and, for comparison, also by scanning electron microscopy (SEM JSM-7600F) image analysis using ImageJ software package.[30]

Transmission electron microscopy (TEM) investigations of unpoled and poled $BiFeO_3$ ceramics were performed on a JEM 2100F, operated at 200 kV and equipped with a JEOL EDXS detector and CCD camera. The specimens were prepared by mechanical grinding, dimpling and final Ar-ion milling or by polishing using tripod polisher to reduce the sample damage associated with conventional ion milling. The peak splitting in the selected area electron diffraction (SAED) patterns were simulated by CrystalMaker program.[31]

For electrical and electromechanical characterization the sintered pellets were thinned to 0.2 mm (for high electric-field measurements) or 0.5 mm (for direct $d_{33}$ measurements), polished and electroded with Au/Cr by sputtering. For the $d_{33}$ measurements the samples were poled by DC electric field of 80 kV/cm (760°C-sintered ceramics) or 50 kV/cm (880°C-sintered ceramics) applied for 15 min at room temperature.

The direct piezoelectric $d_{33}$ coefficient was measured using a dynamic press as a function of static stress ($\sigma_{DC}$) and alternating stress amplitude (expressed here as peak-to-peak value $\sigma$), and frequency.[32] After each measurement the samples were checked for possible depoling; in all the cases, the samples showed the same $d_{33}$ before and after the measurements. The results are presented in terms of $d_0$ ($d_0 = Q_0/F_0$ where $Q_0$ and $F_0$ are amplitudes of charge and force, respectively) and $\tan\delta$ ($\tan\delta = d''/d'$ where $d'$ and $d''$ are real and imaginary components of $d_{33}$, respectively). In some cases the relative $d_0$ is given ($d_0$(relative)), which corresponds to $d_0$(relative) = $d_0(\sigma)/d_0(\sigma=\sigma_{min})$ where $d_0(\sigma=\sigma_{min})$ refers to the $d_0$ measured at the lowest AC stress amplitude ($\sigma_{min}$=0.25 MPa). For samples with $d_{33}$ of about 20 pC/N the values of $d_{33}$ could be measured with an accuracy of about ±0.5 pC/N. The values were mostly affected by the position of the sample in the sample holder. Once the position was fixed, the $d_{33}$ as a function of the amplitude or the frequency of the driving stress was measured with a precision of about ±0.1 pC/N.

Simultaneous polarization–electric-field (P-E) and strain–electric-field (S-E) measurements were performed using an aixACT TF 2000 analyzer equipped with a laser interferometer (aixPES). The measurements were performed by applying to the samples single sinusoidal waveforms of 100 Hz with increasing field amplitude, i.e., 10, 40, 80, 100, 120, 140 and 150 kV/cm. During the measurements, the samples were immersed in silicone oil. S-E curves are plotted by taking the initial strain to be zero.

**III. RESULTS AND DISCUSSION**

**A. Frequency, dynamic and static stress dependence of piezoelectric $d_{33}$ coefficient**

Fig. 1 shows the frequency dependence of $d_0$ and piezoelectric tan$\delta$ of BiFeO$_3$ ceramics sintered at 760°C. In order to explore nonlinear piezoelectric properties, i.e., dependency of the $d_{33}$ on the driving field amplitude, the frequency dispersion of $d_{33}$ was measured at two different stress amplitudes, i.e., 0.6 MPa and 3.2 MPa peak-to-peak. A complex frequency and AC stress dependence of the direct piezoelectric response of BiFeO$_3$ ceramics is revealed by these measurements. Comparison of the frequency dispersion of $d_0$ (Fig. 1a) and tan$\delta$ (Fig. 1b) shows that they are related, as would be expected from Kramers-Krönig relations.[33] For example, in the case of σ=0.6 MPa (Fig. 1a), $d_0$ first increased from 22 pC/N to 26.5 pC/N (i.e., by about 20%) with decreasing frequency from 100 Hz to 0.2 Hz; with further decrease of frequency from 0.2 Hz to 0.04 Hz, the $d_0$ then slightly decreased to 26.3 pC/N, after which it showed again an increase at the lowest frequency range, i.e., below 0.04 Hz, reaching 27.4 pC/N at 0.007 Hz. The same sequence of changes with frequency is observed in the tan$\delta$ (Fig.1b, 0.6 MPa AC peak-to-peak): after an initial increase from 0.025 to 0.085 with decreasing frequency from 100 Hz to 2 Hz, the piezoelectric tan$\delta$ exhibited a decrease as the frequency was lowered reaching a minimum of 0.018 at 0.1 Hz; further decrease of the driving stress frequency below 0.1 Hz again increased the piezoelectric losses up to 0.1 at 0.007 Hz.

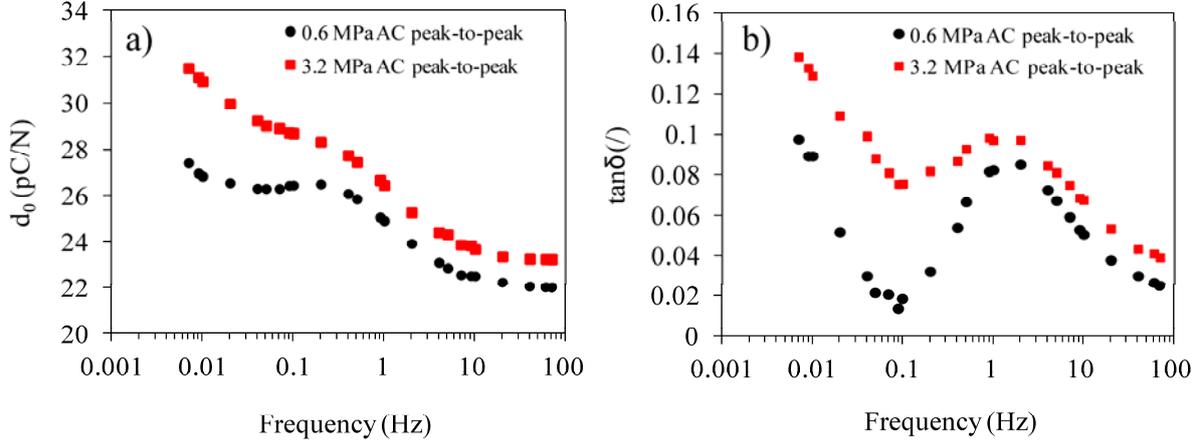

*FIG. 1. a) $d_0$ and b) $\tan\delta$ of $BiFeO_3$ ceramics sintered at 760°C as a function of frequency at 0.6 MPa and 3.2 MPa peak-to-peak field amplitudes. DC stress was set to 3.75 MPa.*

The comparison of $d_0$- and $\tan\delta$-versus-frequency relationships in Fig.1 shows that the changes of $d_0$ and $\tan\delta$ with the driving stress frequency can be described by considering three dispersion processes occurring in different frequency ranges between 100 Hz and 0.007 Hz (see Fig. 1a, 0.6 MPa AC peak-to-peak): i) retardation in the high frequency range (100–0.2 Hz), characterized by an increase of $d_0$ with decreasing frequency, ii) relaxation in the middle frequency range (0.2–0.04 Hz), characterized by a decrease of $d_0$ with decreasing frequency and iii) a second retardation in the low frequency range (<0.04 Hz). Depending on whether $d_0$ increases or decreases with frequency, we refer here to retardation and relaxation processes, although it is common in the literature to refer to both cases as relaxation.[34,35] We shall come back to the possible origin of this complex sequence of relaxational processes in section B.

Increasing the AC driving stress from 0.6 MPa to 3.2 MPa peak-to-peak leads to an increase in $d_0$ (Fig. 1a) and $\tan\delta$ (Fig. 1b) in the whole measured frequency range. As compared to $\sigma=0.6$ MPa, driving the sample with $\sigma=3.2$ MPa leads to a larger increase in $d_0$ with decreasing frequency, particularly below 0.1 Hz (Fig. 1a). In contrast, $\tan\delta$ showed a larger increase with the higher σ at frequencies around 0.1 Hz (Fig. 1b, see the difference between the two curves). In order to get an insight into this nonlinear bahavior, it is interesting to inspect the difference between the $d_0$-versus-frequency (and $tan\delta$-versus-frequency) curves measured at 3.2 MPa and 0.6 MPa, i.e., $\Delta d_0 = d_0(3.2\text{MPa}) - d_0(0.6\text{MPa})$ and $\Delta tan\delta = tan\delta(3.2\text{MPa}) - tan\delta(0.6\text{MPa})$. Assuming superposition of the AC-stress activated contributions,[33] such difference can give information on the origins of the frequency dispersion of this nonlinear, stress dependent part. A typical Debye-like frequency dispersion with a peak in the piezoelectric losses at 0.04 Hz is revealed from this analysis (Fig. 2). Same type of frequency dispersion of the piezoelectric $d_{33}$ coefficient was observed in Sm-doped $PbTiO_3$; however, in contrast to $BiFeO_3$, in the case of the titanate smaller changes were measured in $d_0$ and $tan\delta$ with increasing amplitude of AC stress, possibly suggesting different origins.[25]

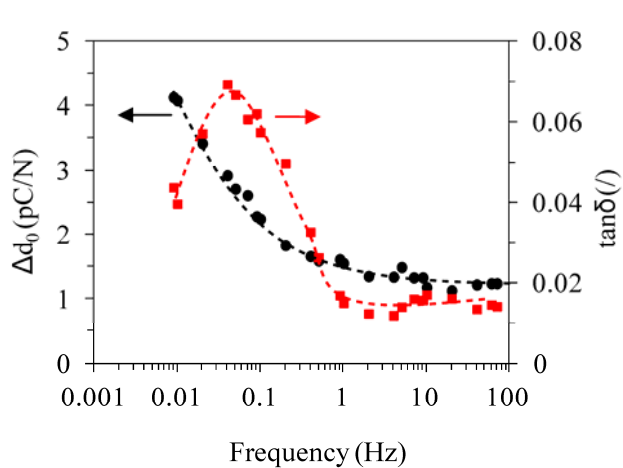

FIG. 2. Difference of $d_0$ and $tan\delta$ curves from Fig.1. The $d_0(tan\delta)$-versus-frequency curve determined at 3.2 MPa of peak-to-peak AC stress was subtracted from the one determined at 0.6 MPa of peak-to-peak AC stress, $\Delta d_0 = d_0(3.2\text{MPa}) - d_0(0.6\text{MPa})$ and $\Delta tan\delta = tan\delta(3.2\text{MPa}) - tan\delta(0.6\text{MPa})$. The lines are drawn as a guide for the eye.

In order to explore the nonlinearity of the direct piezoelectric response of $BiFeO_3$ in more detail, we measured the AC stress dependence of the piezoelectric $d_{33}$ coefficient at selected frequencies. The results of these measurements are presented in Fig. 3. An increase of $d_0$ with AC stress amplitude is evident at all measured frequencies, confirming the nonlinear piezoelectric response (Fig. 3a). A particularity of this nonlinear response consists in the much larger increase of the $d_{33}$ with the stress at 0.01 Hz, as compared to 10, 1 and 0.1 Hz. This can be clearly seen from the plot of the relative $d_0$, which was normalized at the lowest measured AC stress (0.25 MPa), as a function of AC stress amplitude (inset of Fig. 3a). While a relative increase of 8% in $d_0$ with increasing AC stress from 0.25 to 3.2 MPa was measured at 10 and 1 Hz of the driving stress frequency, this relative increase reaches 10% at 0.1 Hz and even 24% at 0.01 Hz (inset of Fig. 3a). The results, therefore, suggest that the nonlinear response of $BiFeO_3$ is strongly dependent on the field frequency, exhibiting increasingly larger contribution to $d_{33}$ with decreasing frequency, particularly below 0.1 Hz. This is in agreement with the assumed frequency dispersion of the nonlinear contribution, shown in Fig. 2, which is characterized by a rise in $d_0$ below 0.1 Hz.

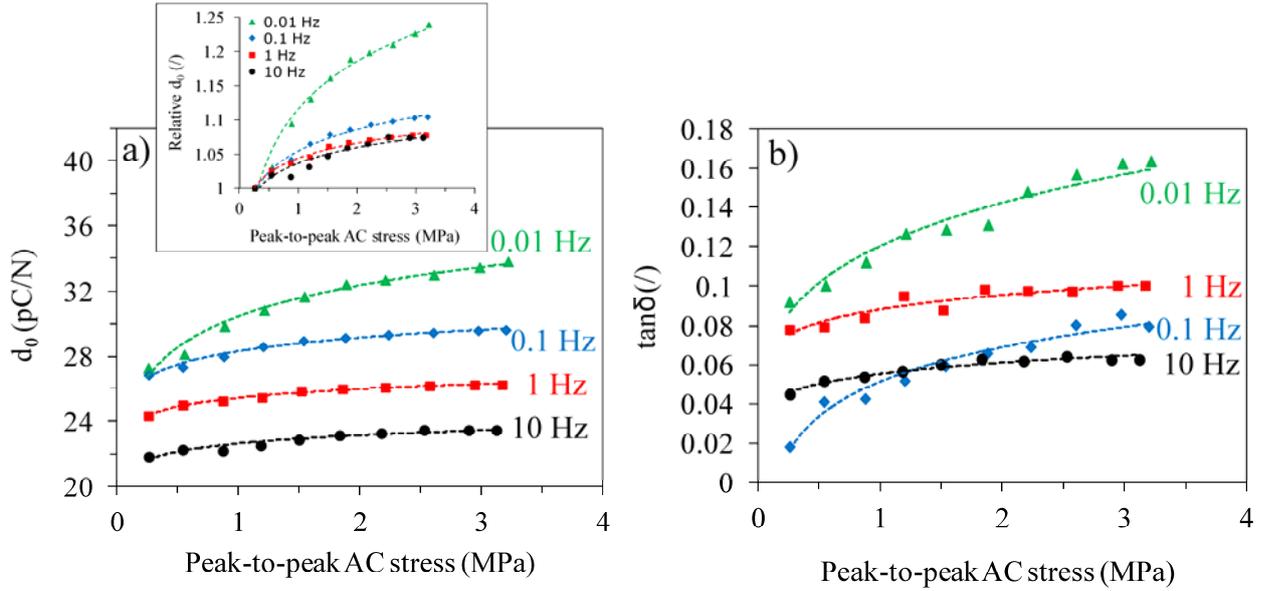

FIG. 3. a) $d_0$ and b) $\tan\delta$ of BiFeO$_3$ ceramics sintered at 760°C as a function of AC stress amplitude at four different frequencies. DC stress was set to 3.75 MPa. The inset of a) shows the relative $d_0$ as a function of AC stress amplitude. The lines are drawn as a guide for the eye.

The process responsible for the piezoelectric nonlinearity in BiFeO$_3$ is also accompanied by energy dissipation, as shown by the increase of tanδ with increasing AC stress amplitude (Fig. 3b). Note that a larger relative increase of tanδ with increasing AC stress amplitude was measured at 0.1 and 0.01 Hz, while a comparably smaller increase of tanδ with AC stress was measured at 1 and 10 Hz; this is again consistent with the frequency dispersion shown in Fig. 2, which predicts large piezoelectric losses close to 0.04 Hz (peak in tanδ), while smaller losses away from the relaxation frequency, i.e., above 1 Hz.

Piezoelectric nonlinearity has been investigated previously in several perovskite and non-perovskite ferroelectric ceramics, including BaTiO$_3$, Pb(Zr,Ti)O$_3$ (PZT) and Bi$_4$Ti$_3$O$_{12}$. The nonlinearity is commonly attributed to extrinsic origins, most commonly to the irreversible movement of non-180° domain walls.[24,25,33,36–39] The ferroelectric-ferroelastic non-180° domains walls, which behave simultaneously as electric and elastic dipoles, can move, both reversibly and irreversibly, under subcoercive external AC electric or stress field, giving a contribution to the measured piezoelectric response. The irreversible non-180° domain wall displacement is responsible for the field dependence of piezoelectric coefficients; for example, the more the amplitude of the AC stress is increased, the more the non-180° domain walls can move irreversibly, resulting in an apparent increase of the piezoelectric coefficient with increasing AC stress amplitude. This irreversible process leads to hysteretic piezoelectric responses as well.

Nonlinear responses are characteristic for materials in which the non-180° domain walls are sufficiently mobile. In contrast, in materials in which the irreversible motion of non-180° domain walls is restricted due to, e.g., clamping by defects or specific domain-wall structure, the piezoelectric coefficient shows only little field dependence or, in some cases, it can even be independent of the field, resulting into linear and non-hysteretic response (if other sources of losses are absent). Examples of polycrystalline ferroelectric materials exhibiting a piezoelectric response with limited or seemingly no extrinsic irreversible contributions over at least weak-to-moderate field range include Fe-doped PZT, Sm-doped $PbTiO_3$, $SrBi_4Ti_4O_{15}$ and Nb-doped $Bi_4Ti_3O_{12}$.[25,33,37,40,41]

According to the common interpretation of the nonlinear piezoelectric response in ferroelectric ceramics in general, the most likely microscopic mechanism responsible for the piezoelectric nonlinearity in $BiFeO_3$ is irreversible motion of non-180° domain-walls. It should be mentioned that, in principle, other moving boundaries, such as phase boundaries, could also lead to similar nonlinear effects. This possibility was discussed in the framework of MPB systems, such as PZT, because near the MPB the free energies of neighboring phases are close to each other, facilitating phase boundary motion under the action of external field.[24,26,42] Crystal phase coexistence has been reported in single phase BiFeO$_3$ thin films with large epitaxial strains.[43] However, in the case of bulk BiFeO$_3$, only a simple R3c rhombohedral structure has been observed and the possibility of interphase boundary motion appears less probable.

We emphasize that the relative increase in the d$_{33}$ of BiFeO$_3$ over the examined AC stress range, which is between 8% (10 Hz) and 24 % (0.01 Hz) (see inset of Fig. 3a), is comparable, relatively, to the nonlinearity in Nb-doped soft PZT with composition close to the MPB.[25,38] This suggests a remarkably large domain-wall contribution to the piezoelectric response of BiFeO$_3$ ceramics. However, we note that the nonlinear piezoelectric response of BiFeO$_3$ exhibits frequency dependence that is distinctly different from that in soft PZT where piezoelectric coefficient decreases linearly with logarithm of the frequency and nonlinearity scales with frequency.[44–46] In BiFeO$_3$, large nonlinear contributions are only observed at low frequencies, i.e., below 0.1 Hz (see inset of Fig. 3a). Possible origins of the nonlinear response of the ferrite are discussed in section E.

Another important aspect of the nonlinear piezoelectric response is its behavior under static stress field. In materials exhibiting large extrinsic domain-wall contribution, e.g., soft Nb-doped PZT (MPB and rhombohedral composition) and BaTiO$_3$, it is commonly observed that the external static compression has an effect of clamping the ferroelastic non-180° domain walls, reducing their irreversible motion; this results in a reduced nonlinear response and reduced hysteresis with increasing compressive stress.[38,39] In contrast, Ochoa et al.[47] recently reported an opposite trend for hard Fe-doped PZT, i.e., increase in the DC bias compressive stress resulted in a larger piezoelectric nonlinearity. This interesting behavior, apparently characteristic for hard PZT materials where the domain walls are strongly clamped by defect complexes, was attributed to a reduced interaction between defects and domain walls by the action of an external DC stress, making subsequent movement of non-180° domain walls easier. Other literature data include tetragonal PZT and fine grained $BaTiO_3$;[38,39] in both cases only a weak effect of the DC stress on the nonlinear response was measured presumably due to the already existing internal stresses in the ceramics related either to the large spontaneous strain in tetragonal PZT (around 3% for the examined PZT43/57 composition) or to the presence of internal stresses in ceramics of $BaTiO_3$ with small grains (< 1 μm).

The nonlinear piezoelectric response of $BiFeO_3$ under different DC stresses (2.1 MPa, 3.4 MPa and 4.5 MPa) is shown in Fig. 4. At all applied DC stresses the piezoelectric coefficient increased with AC stress amplitude; however, as the DC stress was increased, this nonlinear contribution diminished (note the smaller increase of $d_0$ with AC stress when the DC stress field was increased from 2.1 to 3.4 and 4.5 MPa). This behavior is opposite to what was observed in hard PZT by Ochoa et al.,[47] but consistent with what was observed in soft PZT and $BaTiO_3$.[38,39]

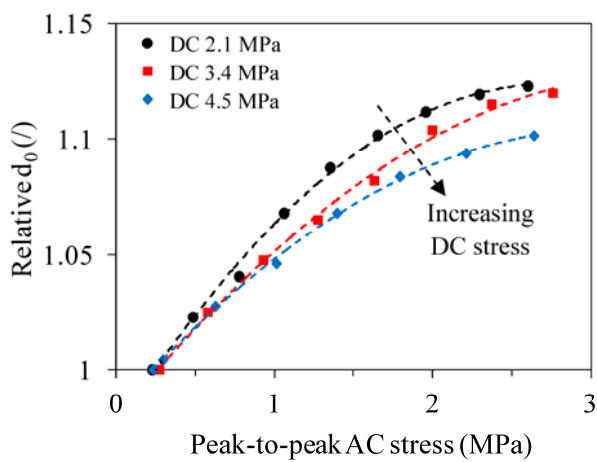

FIG. 4. Relative $d_0$ of $BiFeO_3$ ceramics sintered at 760°C as a function of AC stress amplitude at 0.1 Hz with variable DC stress. The lines are drawn as a guide for the eye.

The fraction of the total $d_{33}$ that is proposed to be due to irreversible displacement of domain walls can be estimated as $(d_{33}(\sigma=\sigma_{max})-d_{33}(\sigma=0))/d_{33}(\sigma=\sigma_{max})$. One obtains 13.2%, 12.6% and 10.8% for DC stresses equal to 2.1 MPa, 3.4 MPa, 4.5 MPa, respectively (see Fig. 4). By comparing with the literature data, we note that, for comparable AC stress field amplitude, the relative decrease of the irreversible contribution, resulting from the increase in the DC stress for $\Delta\sigma_{DC}\approx2.5$ MPa, is smaller in BiFeO$_3$ (from 13.2% to 10.8%) than in BaTiO$_3$ (from 25% to 18%).[38] This probably suggests stronger clamping of domain walls in BiFeO$_3$ compared to BaTiO$_3$.

**B. Linear contribution to the piezoelectric response**

In addition to the nonlinear, AC-stress-activated, component of the piezoelectric response of BiFeO$_3$, another contribution, independent on the AC stress, is revealed from the measurements. A close inspection of Fig. 3b shows that the tan$\delta$ measured at 10, 1 and 0.01 Hz remains large as the AC stress amplitude approaches zero, suggesting an additional lossy contribution to the $d_{33}$, which is independent of the AC stress. Extrapolation of the tan$\delta$ versus AC stress curves to zero AC stress (Fig. 3b) gives tan$\delta$ of 0.038, 0.07, ~0 and 0.065 for frequencies 10, 1, 0.1 and 0.01 Hz, respectively. This tan$\delta$-versus-frequency relationship (not shown separately) resembles the frequency dependence of tan$\delta$ shown in Fig. 1b (see, for example, $\sigma=0.6$ MPa) and suggests an additional linear component of the piezoelectric response with lossy behavior and complex frequency dispersion. As explained in the previous section, this frequency dependence can be described by a sequence of retardation-relaxation-retardation processes in the high (100–0.2 Hz), middle (0.2–0.04 Hz) and low (<0.04 Hz) frequency regions, respectively.

Both relaxation and retardation of the direct piezoelectric $d_{33}$ coefficient and transition between these dispersion regimes with frequency reported here for BiFeO$_3$ (Fig. 1), were also observed in other materials, for example, in Bi$_4$Ti$_3$O$_{12}$.[48] In that case, this frequency dispersion was explained in terms of Maxwell-Wagner piezoelectric relaxation, which has its origin in the coupling between dielectric and piezoelectric properties and may arise in heterogeneous materials, in which the individual components differ in terms of piezoelectric coefficient and electrical conductivity. In the particular case of Bi$_4$Ti$_3$O$_{12}$, this heterogeneity originated from the highly anisotropic grains, which tend to form in the layered Aurivilius phases. The significantly different piezoelectric and conductivity properties along and perpendicular to the growing direction of the grains are responsible for the Maxwell-Wagner relaxation in that material.

BiFeO$_3$ does not possess such anisotropic structure and tendency to form anisotropic grains like Bi$_4$Ti$_3$O$_{12}$; however, it is possible to have similar Maxwell-Wagner piezoelectric effects from other sources. Maxwell-Wagner piezoelectric relaxation could arise wherever uncompensated piezoelectric charges can be created in the material during dynamic stress loading, e.g., at the interphase between the primary material and a secondary phase. If the conductivities of the two phases are different, these piezoelectric charges at the interface will drift with time leading to time-dependent piezoelectric response and, therefore, frequency dependence of the $d_{33}$. In addition to porosity, which was recently discussed in PZT as a possible source of Maxwell-Wagner piezoelectric relaxation,[49] in the case of BiFeO$_3$, one should also consider the often present secondary phases, such as Bi$_{25}$FeO$_{39}$ and Bi$_2$Fe$_4$O$_9$. In fact, depending on the processing conditions, we always observed a small amount (<5%) of these secondary phases in our ceramics.

As porosity and amount of secondary phases in the ceramics depend on the processing conditions, we checked their influence by preparing three different batches of BiFeO$_3$. While the batches denoted here as "batch 1" and "batch 2" were prepared by direct sintering of a mechanochemically activated Bi$_2$O$_3$–Fe$_2$O$_3$ powder mixture, the BiFeO$_3$ from "batch 3" was prepared by first calcining the activated mixture, followed by wet-milling and subsequent sintering. To avoid influences from other parameters, in all the cases, the temperature and time of sintering, i.e., 760°C and 6 h, respectively, were kept the same.

According to XRD, SEM analyses and density measurements, the three batches resulted in BiFeO$_3$ ceramics with different concentrations of secondary phases (Bi$_{25}$FeO$_{39}$ and Bi$_2$Fe$_4$O$_9$), ranging from 1% to 5%, and different porosity, i.e., between 3% and 12%. The different amount of secondary phases in the batch 1 and 2 (1% and 3.3%, respectively), which were prepared following the same procedure, was presumably a result of a lower homogeneity of the initial Bi$_2$O$_3$–Fe$_2$O$_3$ mixture due to more strongly agglomerated initial oxide powders used for batch 2. If the parameters, such as porosity and secondary phases, do not affect or have little influence on the piezoelectric properties through Maxwell-Wagner effect, then the same frequency dispersion would be measured irrespective of the processing conditions. Fig. 5, which shows the piezoelectric tan$\delta$ versus frequency for the ceramics prepared from the three batches, confirms that this is not the case: the tan$\delta$ shows distinctly different frequency dependences for the three batches, particularly in the region 100–0.01 Hz. Note the negative tan$\delta$ of BiFeO$_3$ from batch 2 in the frequency range 0.2–0.5 Hz (Fig. 5, batch 2). Negative piezoelectric phase angle and the corresponding clockwise charge-stress hysteresis, which were also clearly identified in our BiFeO$_3$ ceramics from batch 2, are known to appear in composite piezoelectric materials. Such behavior has been predicted in a serial bilayer model assuming Maxwell-Wagner piezoelectric relaxation and was confirmed experimentally in Aurivilius phases.[48] Therefore, while not directly proving, the present data give support to the possibility of a Maxwell-Wagner mechanism contributing to the piezoelectric response of $BiFeO_3$.

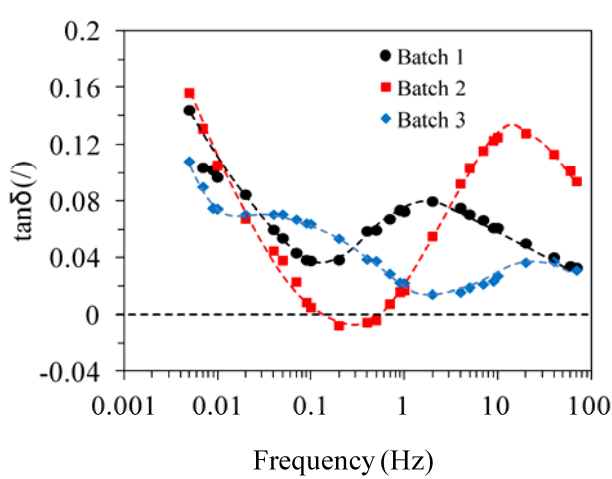

*FIG. 5. Piezoelectric losses (tanδ) as a function of frequency of 760°C-sintered $BiFeO_3$ ceramics prepared from three different batches (batches 1 and 2: direct sintering after mechanochemical activation, batch 3: calcination after mechanochemical activation and subsequent sintering). The peak-to-peak AC stress and DC stress were set to 1.1 MPa and 2.8 MPa, respectively. The lines are drawn as a guide for the eye.*

We emphasize that the nonlinear contributions to $d_{33}$ in the three batches of $BiFeO_3$ were quantitatively similar, i.e., differed by less than 5% of relative $d_0$ at all measured frequencies (not shown). This suggests that the parameters leading to Maxwell-Wagner mechanism are not necessarily coupled to the irreversible domain-wall mobility; they affect primarily the linear, AC stress independent component of the piezoelectric response.

Considering that the concentration of the secondary phases in our samples is low (<5%) and that the strength of the Maxwell-Wagner relaxation depends on the volume fraction of the second phase,[48] we cannot exclude possible "indirect" effects; for example, incomplete reaction between $Bi_2O_3$ and $Fe_2O_3$ and consequent formation of Bi- and Fe-rich secondary phases during processing could result into local regions in the ceramics with non-stoichiometric $BiFeO_3$, which could then, through Maxwell-Wagner mechanism, have a similar effect on the frequency dispersion of $d_{33}$ as secondary phases themselves. That is, the ceramic may consist of three kinds of phases: Bi- and Fe-rich secondary phases, non-stoichiometric $BiFeO_3$, which could not be detected by XRD and SEM, but may occupy a substantial volume of the sample, and stoichiometric $BiFeO_3$.

Another interesting possibility can be considered. A source of Maxwell-Wagner piezoelectric relaxation could be the 71° and 109° domain walls, which were recently shown to exhibit a higher conductivity than the interior of the domains themselves.[50,51] Calculations showed that such conducting domains do not terminate at the electrodes of the sample[52] and, thus, do not form a continuous conductive path through the sample, but would terminate at an insulating layer close to the surface. Conductive domain walls and regions between and around them could lead to a Maxwell-Wagner effect. Moreover, both reversible and irreversible motion of conducting domain walls under applied pressure could affect locally conductivity of grains and could thus, through Maxwell-Wagner effect, influence the macroscopic piezoelectric response. Further systematic studies are necessary to study and elucidate these mechanisms in $BiFeO_3$ in more details.

**C. Effect of poling, quenching and sintering temperature on piezoelectric nonlinearity**

As reported in section A, the piezoelectric response of $BiFeO_3$ is characterized by nonlinearity likely related to the irreversible movement of non-180° domain walls under weak AC stress field. In this section we present further evidences that support the relation between non-180° domain-wall motion, in particular the interaction of the domain walls with the pinning centers, and the nonlinear behavior of $BiFeO_3$.

In our previous study[15] we showed that in $BiFeO_3$ the domain-wall movement, under switching conditions, is restricted due to pinning by defects, probably acceptor–oxygen-vacancy defect complexes. Macroscopically, this is manifested as pinched and biased P-E loops, as illustrated in Fig. 6a (full line). We also showed that depinching, i.e., opening of the P-E loop and the corresponding increase of the remanent polarization, could be achieved by means of three methods: i) electric-field cycling, ii) rapid quenching from above the Curie temperature ($T_c$) and iii) high-temperature annealing (>820°C). We assume here that if these processes affected high-field properties of $BiFeO_3$, by facilitating the movement of domain walls over large scale (switching), they might affect weak-field piezoelectric properties, through enhanced small-scale domain-wall motion at subswitching conditions. We next analyze domain-wall depinning achieved by the three aforementioned methods.

We first consider depinning of domain walls by electric-field cycling.[15,23] Exposing the material to AC field cycling results into a reduced interaction between defects and domains walls, i.e., domain walls are released from pinning centers as a consequence of the rearrangement of defects.[53] Considering that this defect rearrangement occurs under the action of an external alternating electric field, we may consider that similar depinning could also occur during poling of the ceramics. Poling of BiFeO$_3$ was conducted under high DC electric field, i.e., 80 kV/cm; the long exposure of the ceramics under this field, e.g., 15 minutes, could allow kinetically a rearrangement of defects by diffusion and, consequently, a reduced pinning effect.

To verify this hypothesis we compare P-E and S-E hysteresis loops of BiFeO$_3$ before and after poling (Fig. 6). The sample before poling refers to as the virgin (as-sintered) sample. Opening and depinching of both P-E and S-E loops is clearly observed after the material was exposed to DC poling (compare loops of non-poled and poled samples in Fig. 6). Note the larger remanent polarization (P$_r$) and peak-to-peak strain (S$_{pp}$) of the poled ceramics (2P$_r$ = 40 µC/cm$^2$, S$_{pp}$ = 0.16%; Fig. 6, dashed curves) as compared to the non-poled ceramics (2P$_r$ = 17 µC/cm$^2$, S$_{pp}$ = 0.068%; Fig. 6, full curves).

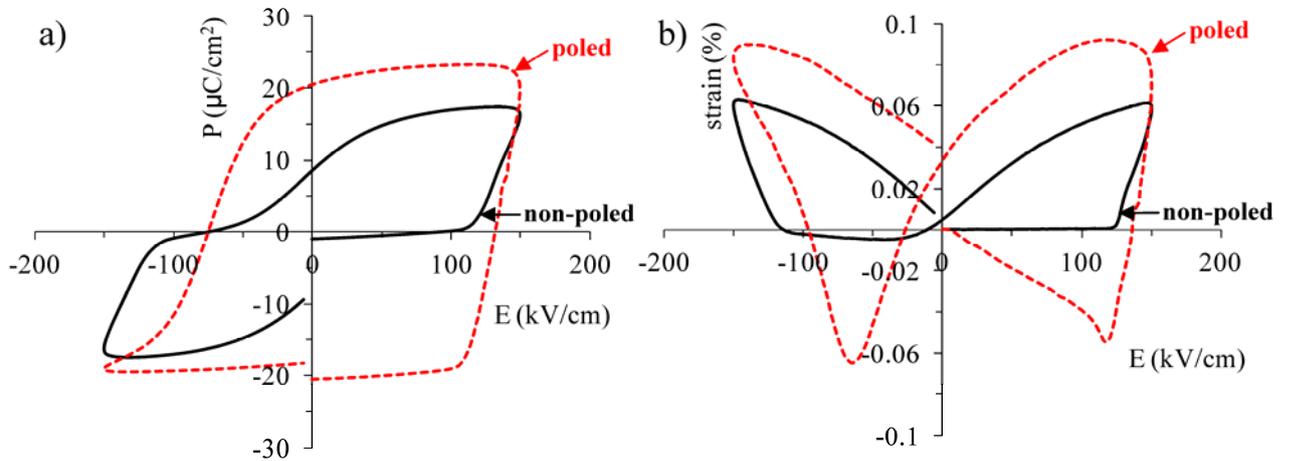

FIG. 6. a) P-E and b) S-E hysteresis loops of BiFeO$_3$ ceramics sintered at 760°C before (non-poled) and after poling at 80 kV/cm of DC electric field (poled).

In the poled sample (Fig. 6b, dashed curve) the strain initially decreases until reaching a minimum value of -0.055% at 120 kV/cm; the slope of this linear portion of the S-E curve is 40 pm/V, which is consistent with the $d_{33}$ of $BiFeO_3$.[3,15] Therefore, this decrease of strain confirms the poled state of the sample, i.e., the sample initially shrinks as the electric field is applied against poling direction. It should be noted that this poled state is a consequence of both poling field, in which we used 80 kV/cm of DC field, and re-poling of the sample by application of the AC field during polarization/strain measurements, because several AC cycles with increasing amplitudes were applied before the final one with amplitude of 150 kV/cm, shown in Fig. 6 (see also description in experimental part). After reaching a minimum, the strain of the poled sample showed an abrupt rise above the coercive field of 120 kV/cm, which is due to domain switching. Same changes in the strain are repeated at the field with negative polarity.

The strain response of the non-poled sample is in striking contrast with the one that was poled by a DC field of 80 kV/cm (Fig. 6b, full curve). In the non-poled sample, firstly, no measurable strain can be detected until 120 kV/cm, which means that, in contrast to the poled sample (Fig. 6b, dashed curve), previously applied AC cycles with lower amplitude (<150 kV/cm) were insufficient to pole the material. The large strain response of the poled sample below coercive field (<120 kV/cm for positive field polarity) contrasts the nearly zero strain of the non-poled sample, confirming the higher electromechanical response of the poled sample at weak electric fields (compare strain of non-poled and poled samples at low fields in Fig. 6b). Secondly, a strong restoring force, reflecting domain-wall pinning, is observed in the non-poled sample upon releasing the field from 150 to 0 kV/cm, resulting in a nearly zero remanent strain (see strain at zero field in Fig. 6b, full curve). Thus, comparison of the electric-field induced strain in non-poled and poled ceramics in Fig. 6b confirms a larger domain-wall movement and switching after the material was exposed to DC poling. Similar effects have been earlier discussed in refs. 33,54 and 55.

The observed depinning of domain walls by exposing $BiFeO_3$ to high DC electric field (poling) might explain the apparent discrepancy between the rather large relative nonlinearity and "soft" behavior of $BiFeO_3$, revealed from the present piezoelectric measurements, and a more "hard" behavior reported earlier in ref. 15 which was inferred, however, from the experiments on as-sintered, non-poled ceramics.

The second method that we used to depin domain walls from defects in $BiFeO_3$ is quenching. The opening of the P-E loop after quenching the ferrite from above $T_c$ into water, reported in ref. 15, is a consequence of the increased domain wall mobility due to freezing of the disordered state of the defect complexes. The defect complexes in disordered state, characteristic for the paraelectric phase, are less effective in pinning the walls than the ordered complexes, characteristic for the ferroelectric phase in the aged state; the disordered defect state can be frozen in the ferroelectric phase by rapid cooling of the ceramics from above $T_c$.[56]

If domain walls are released from pinning centers by quenching, then a larger contribution of these walls to the piezoelectric response is expected not only at switching conditions but also at weak-to-moderate driving fields. Fig. 7 shows that this is indeed the case. Clearly, the quenched sample exhibits larger nonlinearity (35% of increase in $d_0$ at maximum $\sigma$), compared to the non-quenched sample (24% of increase in $d_0$ at maximum $\sigma$). This is consistent with the higher non-180° domain-wall mobility after the sample was quenched. It should be noted, however, that this larger nonlinear contribution in the quenched sample was only found at the lowest frequency, i.e., 0.01 Hz (the same degree of nonlinearity was measured in both quenched and non-quenched samples at 0.1 Hz). This suggests that the frequency dependence of the nonlinear contribution to $d_{33}$ in $BiFeO_3$ (see inset of Fig.3a and section E) is preserved after quenching and might be associated with a mechanism, which is little affected by quenching or defect disorder. In addition, a partial realignment of the quenched (disordered) defects may occur during poling; this could alter the interaction between defects and domain walls. It is also possible, for example, that partially ordered defects are disordered by the low frequency field, such as at 0.01 Hz, as shown to be the case in hard PZT.[54]

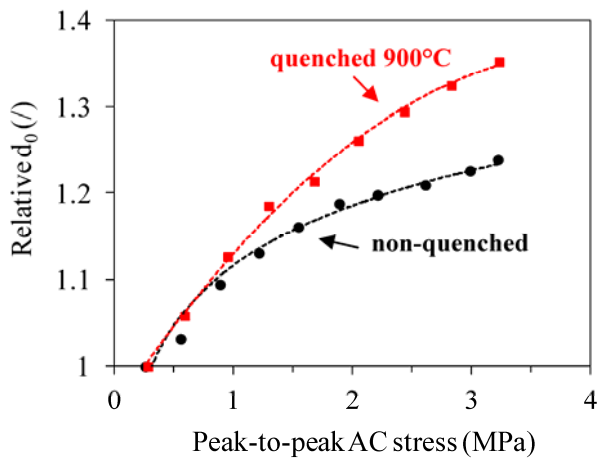

FIG. 7. Relative $d_0$ of non-quenched and 900°C-quenched BiFeO$_3$ ceramics sintered at 760°C as a function of AC stress amplitude at 0.01 Hz. DC stress was set to 3.75 MPa. The lines are drawn as a guide for the eye.

The third method used for domain-wall depinning is annealing BiFeO$_3$ at elevated temperatures. Apparently, the larger domain-wall mobility after annealing at high temperatures (>820°C) has its origin in the creation of defects, most probably bismuth-vacancy–oxygen-vacancy defect pairs, which result from the sublimation of the volatile Bi$_2$O$_3$.[15] This result suggests that the presence of oxygen vacancies is not a sufficient condition for significant domain wall pinning.

Fig. 8 compares the P-E and S-E loops of BiFeO$_3$ sintered at 760°C and 880°C. In agreement with our previous study, opening of the P-E loop and the corresponding increase of the remanent polarization and decrease of the coercive field were observed if the ceramics were sintered at 880°C (2P$_r$ = 40 μC/cm$^2$, E$_c$ = 65 kV/cm) as compared to 760°C (2P$_r$ = 17 μC/cm$^2$, E$_c$ = 92 kV/cm) (Fig. 8a). An analogous depinching effect due to high-temperature sintering is seen in the S-E loop (Fig. 8b); in fact, the 880°C-sintered sample shows larger peak-to-peak strain (S$_{pp}$ = 0.13%) in comparison with the 760°C-sintered sample (S$_{pp}$ = 0.068%).

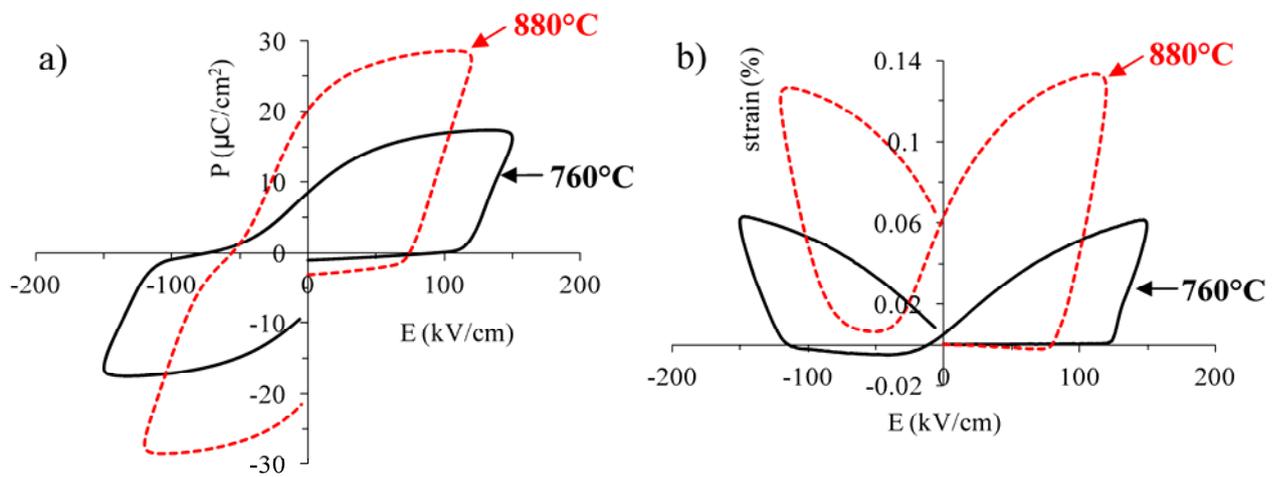

FIG. 8. a) P-E and b) S-E hysteresis loops of BiFeO$_3$ ceramics sintered at 760°C and 880°C.

It has recently been shown for A- and B-site acceptor doped BiFeO$_3$ films (Ca$^{2+}$- and Ni$^{2+}$-doped, respectively) that the domain-wall pinning is significantly stronger in the film with B-site acceptor–oxygen vacancy defect complexes than in the one with A-site acceptor–oxygen vacancy complexes.[57] Thus, it is possible that the depinning effect, which we observed in the case of BiFeO$_3$ annealed at high temperature, could be related to a transition between stronger pinning due to, e.g., Fe$^{2+}$–V$_O^{\bullet\bullet}$ centers, present in the low-temperature sintered ceramics, to weaker pinning, e.g., V$_{Bi}^{'''}$–V$_O^{\bullet\bullet}$ centers, which could be created during sintering at elevated temperatures by sublimation of Bi$_2$O$_3$. In fact, the P-E loop of the 880°C-sintered sample is still pinched (Fig. 8a), however, to a lesser extent than the loop of the 760°-sintered ceramics; such difference in loop pinching was also found by comparing the P-E loops of the two acceptor-doped BiFeO$_3$ films.[57]

If larger electric-field-induced strain, mostly coming from non-180° domain-wall movement,[23] is observed in the 880°C-sintered ceramics (Fig. 8b, 880°C), then, again, we can expect a larger contribution of these domain walls to the d$_{33}$ measured at weak fields. The result from Fig. 9, showing the five-times larger nonlinear contribution to the d$_{33}$ in the 880°C-sintered ceramics as compared to the 760°C-sintered sample (at maximum AC stress), is consistent with this hypothesis. Interestingly, like in the case of the quenched sample, we observed the nonlinear response at low frequencies also in the 880°C-sintered sample (not shown). Finally, we emphasize the notably larger irreversible contribution of non-180° domains walls in the high-temperature sintered sample, which reaches 38% of the total d$_{33}$. The results confirm the importance of controlling the sintering conditions as they affect considerably the piezoelectric response, particularly at higher AC stress levels.

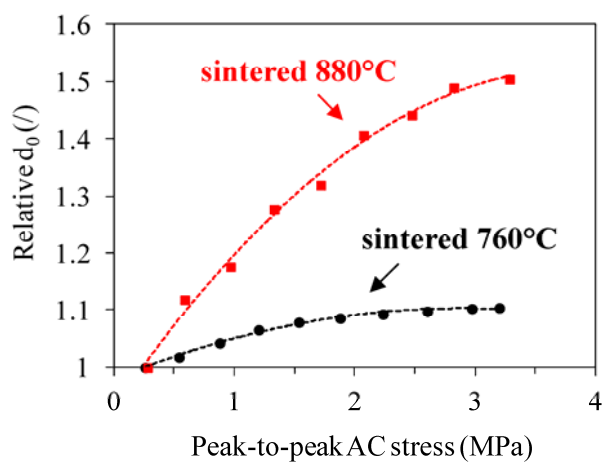

FIG. 9. Relative d$_0$ of BiFeO$_3$ ceramics sintered at 760°C and 880°C as a function of AC stress amplitude at 0.1 Hz. DC stress was set to 3.75 MPa. The lines are drawn as a guide for the eye.

In addition, we note that the relative density of the 880°C-sintered sample (95%) was higher than that of the 760°C-sintered sample (92%), which could affect the piezoelectric nonlinearity. To verify this, we prepared another set of samples, which were calcined and milled before being actually sintered at 760°C (this batch is denoted as "batch 3" in Fig. 5). These calcined and sintered samples resulted in higher density, i.e., 97% relative. Within 5% difference in the measured relative $d_0$, the piezoelectric nonlinearity of the 97%-dense ceramics was comparable to the 92%-dense ceramics (not shown). This probably rules out density as being the primary parameter leading to enhanced piezoelectric nonlinearity, as shown in Fig. 9.

**D. Domain structure of BiFeO$_3$ determined by TEM**

Since the motion of non-180° domain walls is responsible for a considerable part of the piezoelectric response of BiFeO$_3$ both at strong and weak fields, it is of interest to have a deeper look at the domain structure of BiFeO$_3$ ceramics.

A bright-field (BF)-TEM image of typical 50-nm-sized domains in a BiFeO$_3$ grain is shown in Fig. 10a. In addition, we found characteristic defects in the ceramics, shown in Fig. 10b, which were identified as antiphase boundaries (APBs). The corresponding SEAD pattern of such region is shown as inset of Fig. 10b. Here, we identified superlattice reflections at 1/2{hkl} positions (marked in the inset of Fig. 10b). As unambiguously demonstrated by Woodward et al.[58], these superlattice reflections are due to the doubling of the unit cell related to anti-phase rotation of FeO$_6$ octahedra around the rhombohedral (111) axis, i.e., octahedra tilting.

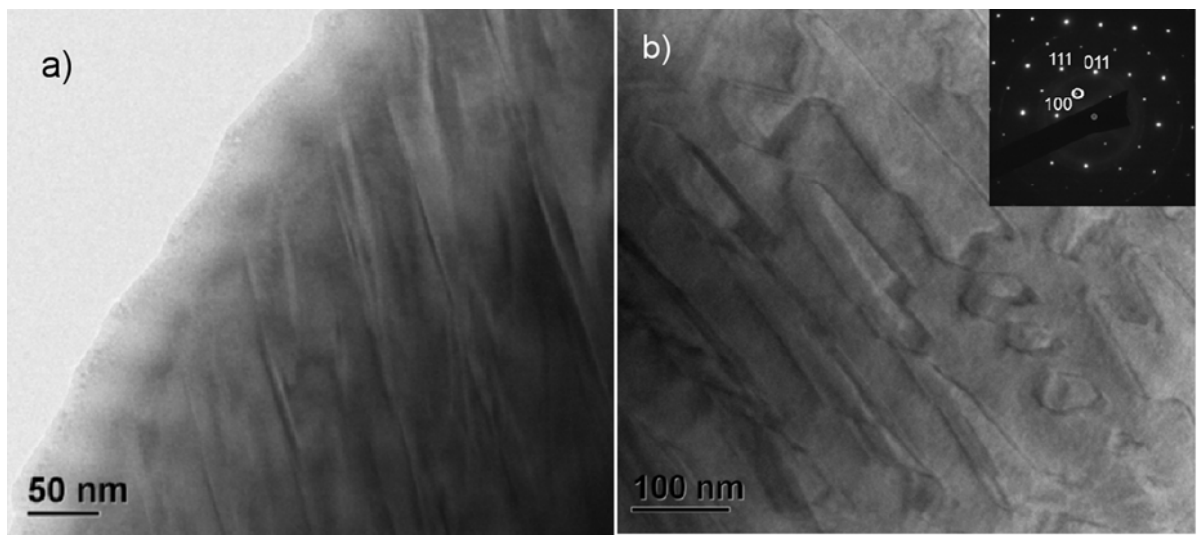

*FIG. 10: a) TEM-BF image of a typical ~50 nm sized domains and b) image of antiphase boundaries in a grain of 760°C-sintered BiFeO$_3$ ceramics in [01-1] zone axis. The inset of b) shows the SAED pattern taken from the sample region shown in b). The marked diffraction spot on the inset corresponds to the set of spots in the (½ ½ ½) positions, which are due to oxygen shifts in the octahedrally tilted BiFeO$_3$ structure.*

In the rhombohedral BiFeO$_3$ (space group R3c, ICSD 15299) three different types of domain walls are possible: 71°, 109° and 180° domain walls.[3] In order to identify the type of the domain in our BiFeO$_3$ ceramics, SAED patterns were taken from different areas in the sample. The SAED patterns of the grains oriented in [110] zone axes, taken on the domains from an area of ~700x700 nm$^2$, showed splitting of the {111} reflections (see Fig. 11A and 11B, marked spots). Such patterns can be experimentally observed only in the case where the polarization vectors are inclined to each other by an angle of 71° and/or 109°, whereas in the case of 180° domains there is no splitting. Next, we performed a simulation of peak splitting for both types of domains; the simulated SAED patterns for 109° and 71° domain walls are shown in Figs. 11a and 11b, respectively. For the simulations we used a pseudo-cubic unit cell with the cell parameter 3.965 Å and rhombohedral angle of 89.3°.[3] The experimentally observed splitting of {111} reflections (Fig. 11A and 11B) can be clearly explained by the presence of 109° and 71° domains (compare experimental SAED from Fig. 11A and 11B with simulated SAED from Fig. 11a and 11b, respectively).

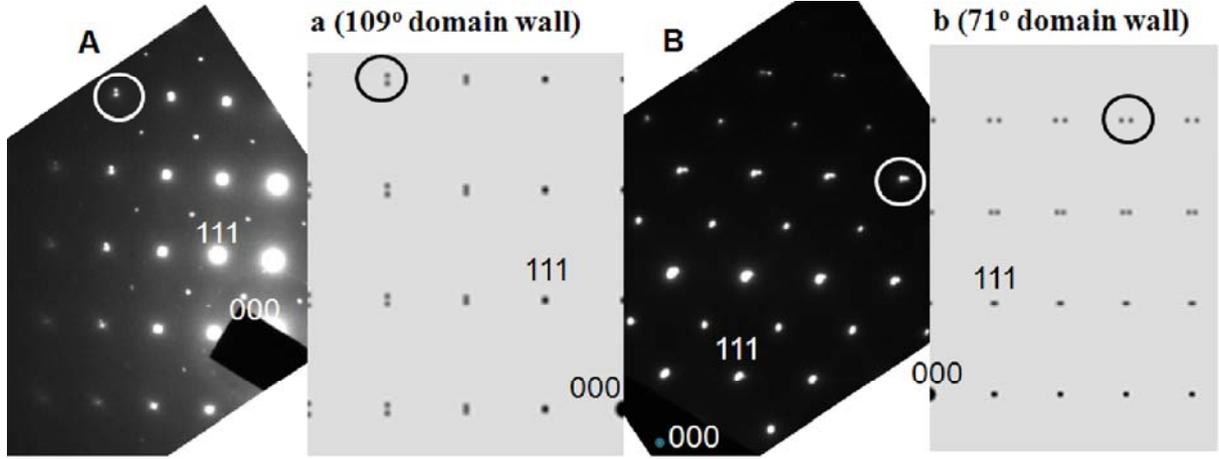

FIG. 11: A) and B) Experimental SAED patterns from domains in a 760°-sintered BiFeO$_3$ ceramics in <110> zone axes where splitting of the marked (333) reflections was observed in approximately A) <110> and B) <100> directions. For simulation of SAED spot splitting we used a) one domain in [0-11] and another in [01-1] zone axes (109° domain) and b) one domain in [10-1] and another in [-101] zone axes (71° domain).

We note that the splitting characteristic for 71° domain walls (Fig. 11B) was mostly seen in unpoled BiFeO$_3$, whereas in poled BiFeO$_3$ ceramics, beside 71° domains, the splitting characteristic for 109° domain walls was also observed (Fig. 11A). Apparently, the majority of 109° domain walls nucleated during poling. For comparison, a higher population of 71° than 109° domain walls was also observed in non-poled rhombohedral PZT.[59]

### E. Discussion on piezoelectric nonlinearity in BiFeO$_3$

Ferroelectric oxides with strong nonlinearity in dielectric and piezoelectric properties tend to exhibit a broad frequency dispersion of these properties, where the permittivity or piezoelectric coefficient increase logarithmically with decreasing frequency ω.[44–46] For example, in PZT thin films the permittivity obeys a log(1/ω) dependence over a frequency range of six orders of magnitude, i.e., from 0.01 to $10^4$ Hz.[33] Same type of frequency dispersion was also found in the piezoelectric $d_{33}$ coefficient of Nb-doped soft PZT and Nb-doped $Bi_4Ti_3O_{12}$ ceramics. Consistent with this dispersion, the piezoelectric nonlinear parameters in soft PZT ceramics exhibited the same frequency dependence. This indeed confirmed that nonlinearity and frequency dispersion are coupled and are related to displacement of domain walls, which in these nonlinear systems constitute a great part of the total electromechanical response.[44,46]

The linear increase of the $d_{33}$ with decreasing log(ω), such as observed in soft PZT and other ferroelectrics,[25,46] was not observed in $BiFeO_3$. Instead, the nonlinearity of the ferrite exhibits a different frequency dependence, which is characterized by large increase in the nonlinear response only at low frequencies, i.e., below 0.1 Hz (see Fig. 2 and inset of Fig. 3a). If this low-frequency nonlinearity would be related to a domain-wall depinning upon application of dynamic stress of longer period, then one would expect this behavior to be altered, possibly changing to a different regime, by releasing domain walls from the pinning centers. Interestingly, this low-frequency nonlinearity is preserved after both quenching and high-temperature sintering, in spite of the fact that both methods provided depinning of the walls (see section C). In the following, we propose possible interpretations of the frequency dependence of the nonlinear behavior of $BiFeO_3$ ceramics.

One possibility for the distinct frequency dependence of the nonlinear response of $BiFeO_3$ ceramics would be the presence of octahedra tilting. It was shown, for example, that in $Pb(Zr_xTi_{1-x})O_3$ compositions belonging to the R3c space group (x = 0.63–0.90) with octahedra tilt angle between 0° and 6° the domain walls are effectively pinned by the pattern of rotational tilts.[60] It was proposed that the long-range structural distortion related to the octahedra tilts increases the strain energy of domain walls, resulting in a higher activation energy for domain wall motion.[61] This resulted in a lower nonlinear contribution from domain walls to $d_{33}$ in the tilted R3c than in the parent non-tilted R3m phase.[62] No data are available on the frequency dependence of the direct piezoelectric response in these PZT compositions; it could be particularly interesting, therefore, to explore if depinning of domain walls in such octahedrally tilted structures can be accomplished by applying low frequency dynamic stress. We note that $BiFeO_3$ has the same R3c tilted structure as PZT, however, the tilt angle is quite larger than in PZT, i.e., 11–14°,[1,63] and one could expect a stronger pinning. The results of recent experiments on PZT, which will be published separately, indicate similar piezoelectric dispersion in R3c and R3m PZT and may rule out tilting as a possible origin of the low-frequency nonlinearity as observed in $BiFeO_3$.

Recently, it was suggested that, due to screening of the polarization discontinuity arising from a non-zero normal component of polarization across a 109° domain wall in $BiFeO_3$, charge can accumulate at these 109° domain walls.[50] In addition, formation of head-to-head or tail-to-tail domain-wall configurations, as discussed by Maksymovych et al.,[64] could also lead to accumulation of compensating charges at the wall. The formation of charged domains walls (head-to-head or tail-to-tail) was recently demonstrated by piezoforce microscopy (PFM) during switching of a $(111)_{pc}$ oriented $BiFeO_3$ thin film. In fact, the 180° domain-wall reversal in $BiFeO_3$ occurs via intermediate non-180° domain switching (multistep switching mechanism), during which non-neutral domain wall configurations appeared and stabilized.[65] We note that our recent results on the electric-field induced strain in $BiFeO_3$ ceramics agree with the multistep switching mechanism.[23] This suggests a possibility of having charged domain walls in BiFeO$_3$ ceramics, which could nucleate, for example, during the poling procedure when large electric field are applied to the material.

Charge accumulated at the domain wall area might reduce the mobility of domain walls or lead to a "creep-like" wall motion as displacement of a domain wall upon the action of external field now requires charge migration.[53,66,67] Eventually, a depinning effect could arise by applying on a material AC stress (or electric) field for a sufficiently long time (low frequency) so as to allow the diffusive displacement of charges.[66] If this is the case, one might expect a strong influence of the frequency of the driving stress on the nonlinear response, which was effectively measured in our BiFeO$_3$ ceramics.

In addition to the possibilities described above, one should also consider possible coupling between piezoelectric nonlinearity and its frequency dispersion mediated by electrical conductivity. Local measurements by AFM in thin BiFeO$_3$ films recently revealed that both 71° and 109° domain walls in BiFeO$_3$ are more conductive than the domain itself.[50,51] Displacing irreversibly a conductive non-180° domain wall with an external AC stress can alter the conductive path through these walls, which might affect, macroscopically, the frequency dispersion of the piezoelectric response through Maxwell-Wagner piezoelectric effect. Recent simultaneous measurements of conductivity and converse piezoelectric response by means of AFM showed indeed coupling between local displacement and conductivity of domain walls in BiFeO$_3$ thin film.[64] The macroscopic manifestation of such mechanism remains an open question, but present experiments may reflect coupling of conducting domains walls and piezoelectric dispersion.

We recently showed that cracks can appear in BiFeO$_3$ ceramics during poling when a certain threshold DC field (approximately 90 kV/cm for 760°C-sintered ceramics), was exceeded.[15] The origin of the cracks is most probably the large strain experienced by the grains during non-180° domain switching.[23] Even if macroscopic cracks in the samples used in this study, which were intentionally poled below the threshold DC field, were not observed, there is still a possibility of the presence of microcracks. Since the crack surface could act as a source of uncompensated piezoelectric charges created during AC stress loading, microcracks could possibly influence the frequency dispersion of the $d_{33}$ via Maxwell-Wagner relaxation, as recently discussed for anisotropic pores in PZT.[49] In addition, they could even play a more active role, for example, by opening and closing under external AC stress field, which could also affect the piezoelectric nonlinearity. However, our recent SEM investigations (not reported here) did not show any evidence of the presence of microcracks in poled samples.

## IV. SUMMARY

Measurements of the frequency, dynamic and static stress dependence of the direct piezoelectric $d_{33}$ coefficient revealed several unusual features of the piezoelectric response of the BiFeO$_3$ ceramics. They can be summarized as follows:

1.) Large piezoelectric nonlinearity, characterized by a relative increase in $d_0$ of up to 50% with increasing AC stress from 0.2 to 3.2 MPa peak-to-peak, was measured at low frequencies. This nonlinear response has a frequency dependence, which is atypical with respect to the general $d_{33}$ frequency dispersion measured in other piezoelectrically nonlinear ferroelectric ceramics, such as, for example, soft PZT.

2.) The nonlinear response can be attributed to irreversible non-180° domain wall movement under AC driving pressure. Both 71° and 109° ferroelectric-feroelastic domain walls were identified in poled $BiFeO_3$ ceramics using TEM analysis.

3.) A correlation was found between large-signal (polarization– and strain–electric-field loops) and small-signal (direct piezoelectric) properties. Deppining of domain walls by means of either quenching or high-temperature sintering, which resulted in higher domain-wall mobility at switching conditions, increased the weak-field nonlinear piezoelectric response as well. The results are consistent with conclusion 2.) and with the general observation in ferroelectric ceramics where piezoelectric nonlinearity is usually attributed to extrinsic domain-wall contributions.

4.) The linear, stress independent, contribution to $d_{33}$ in $BiFeO_3$ is characterized by a lossy behavior, complex frequency dispersion and, depending on the processing conditions, also by negative piezoelectric losses (clockwise charge-stress hysteresis loop), which might suggest Maxwell-Wagner piezoelectric relaxation.

**ACKNOWLEDGEMENTS**

The work was carried out within the Research Program "Electronic Ceramics, Nano, 2D and 3D Structures" P2-0105 of the Slovenian Research Agency. The author thanks personally Prof. Dr. Nava Setter for giving financial support for this work.

We give special thanks to Tanja Urh, Maja Hromec and Larisa Suhodolčan for the preparation of the samples. Edi Krajnc is gratefully acknowledged for the XRD analyses. Special thanks go to Brigita Kužnik for SEM analyses and preparation of the samples for TEM.